\documentclass[fleqn,10pt]{wlscirep}
\title{Detecting the impact of public transit on the transmission of epidemics}

\author[1,*]{Zhanwei Du}
\author[1]{Yuan Bai}
\affil[1]{Jilin University, Changchun, Jilin, 130012, China}

\affil[*]{duzhanwei0@gmail.com}

\begin{abstract}
In many developing countries, public transit plays an important role
in daily life. However, few existing methods have considered
the influence of public transit in their models. In this work, we
present a dual-perspective view of the epidemic spreading
process of the individual that involves both contamination in places (such as work
places and homes) and public transit (such as buses and trains). In
more detail, we consider a group of individuals who travel to some
places using public transit, and introduce public transit into the
epidemic spreading process. A novel modeling framework is proposed
considering place-based infections and the public-transit-based infections. In the urban scenario, we investigate the public transit trip contribution rate (PTTCR) in the
epidemic spreading process of the individual, and assess the impact of the public
transit trip contribution rate by evaluating the volume of infectious
people. Scenarios for strategies such as public transit and school
closure were tested and analyzed. Our simulation results suggest that
individuals with a high public transit trip contribution rate will
increase the volume of infectious people when an infectious disease
outbreak occurs by affecting the social network through the public transit trip contribution rate.

\end{abstract}
\begin{document}

\flushbottom
\maketitle
\thispagestyle{empty}


\section*{Introduction}
Epidemiological research is complex and involves aspects such as public
policies~\cite{bib1}. There are an increasing number of applications and tools that support real-world decisions in public health epidemiology~\cite{bib16,bib17}.
It is critical to use mathematical models to analyze epidemic spreading 
in public health epidemiology, to help overcome the problems of sparse observations, inference with missing data, and so on~\cite{bib1,bib2}.
These models can describe the spreading mechanisms of viruses,
quantify the interventions' effects, and identify key factors related to the course of an outbreak\cite{bib2}. 
 Infectious disease spreading models are powerful tools for controlling
 the development of infectious diseases ~\cite{bib2}. Mathematical and
 computational models have proven useful when addressing the 2014 Ebola outbreak, the 2009 H1N1 outbreak, and so on.

Consider the difference between urban transport in China and other
Western countries; people would like to choose public transit for
their daily travel needs \cite{bib18}. For example, the average public
transit trip contribution rate is approximately 9\% ~\cite{bib3} in
the United States. In contrast, in some cities in China, such as
Beijing and Shanghai, the average public transit trip contribution
rate can be as high as 40\%. However, individual-based models are
important tools for studying the transmission mechanism of pandemic
influenza ~\cite{bib4}. With regards to individual-based models,
many of these models ignore the possible risk inherent in commuting. For example, a heterogeneous graph modeling method was used by Dongmin et al. to describe the dynamic process of influenza virus transmission using clinical data  ~\cite{bib5}.
Meyers et al. attempted to model the early epidemiology of SARS. They
applied contact network epidemiology to show that two outbreaks may
have clearly different behaviors~\cite{bib6}. FluCaster, a disease
surveillance tool, can track and predict the process of disease spreading~\cite{bib7}.
The VirusTracker app simulates the spread of a virus and highlights
the critical role of vaccinations in combating a disease
outbreak~\cite{bib9}. Dynamic Behavior Visualizer is a tool for visualizing people's dynamic behaviors and movements in a disaster~\cite{bib10}.
However, these aforementioned models primarily assume that people
are infected at a fixed location. Public transit, as a high-risk
place where different crowds contact each other, is often
neglected. This overlook can be understood, particularly in some
developed countries (such as the United States) where almost ninety
percent of the people travel by private vehicles and few travel by public
transit or other modes of public transportation~\cite{bib3}. These methods
cannot describe the infection risk due to crowded public transit, such as
that in large cities in China. Thus, we can see that the intervention of public transit can affect the trend of epidemic diseases to some extent and enlarge the transmissibility threshold to some extent (see the analysis of the transmissibility threshold in the Methods section). 
In this regard, a new model is needed that considers urban public transit. In this respect, our work may provide a means for modeling the impact of public transit trips and for estimating the effectiveness of infection controls during public transit trips~\cite{bib11}. 

\section*{Materials and Methods}
We use the data of a resident in Blacksburg from
http://ndssl.vbi.vt.edu/synthetic-data/ to construct the social
network. We know when and where the resident is for one
day. Moreover, the public transit information data can be obtained from
http://www.gtfs-data-exchange.com/agency/blacksburg-transit/. The
detailed Supporting Information that we use is described in
Table~\ref{Tab:table1}. With the public transit information, we can
schedule people’s daily trips with the help of Google Maps. The
resident information includes resident identification, start time of
resident activity, resident location, the duration in this location,
and so on. The public transit information includes public transit
stops, trips, stop times, routes, and so on. Then, we propose a novel
modeling framework for describing the dynamic process of individuals
transmitting influenza virus by integrating public transit. In
this model, we use the chain-binomial model in
susceptible-exposed-infected-recovered (SEIR) models based
on~\cite{bib12} to simulate the disease transmission process in public
transit. We use this model to identify chains of transmission and to
estimate the probability of an epidemic with the time measured in
arbitrary discrete units. The order of transmission events is
arbitrary. The most important feature is that our modeling approach was combined with Google Maps to schedule the daily public transit travel of an individual, 
analyze the agent behavior under different scenarios, and simulate the interventions. More details of the above methods can be found in the Supporting Information.

\begin{table}[h]
\centering
\caption{\label{Tab:table1}Supporting Information.}
\begin{tabular}{|p{4.0cm}|p{3.0cm}|} \hline 
Resident Information&Transit Information \\ \hline 
PID&trips\\  \hline 
STRATTIME&stops\\ \hline 
DURATION&stop\_times\\ \hline 
LOCATION&routes \\ \hline 
\end{tabular}
\end{table}

\section*{Results}
We conduct the simulation of the epidemic spread in Blacksburg, Virginia, USA (see Methods), using our public-transit-based modeling framework (PTF) approach. 
This approach has two layers. The first primarily captures the synthesized
population's mobility, featured by the population's geographic
heterogeneities. The second layer focuses on an individual-level model
that is embedded in the PTF. With this model, we can describe the
 mechanism of the virus in interactions between individuals. People were assigned
to homes, schools, workplaces, and other public places depending on
their individual properties~\cite{bib13}. A day consists of different
schedules. An individual’s contacts can occur at the above four
places. Here, individuals can make contact with all of their family
members. Additionally, when people are in different places, they can make contact with others in the same places. Due to the activity generation method~\cite{bib14}, there is no change in the daily mobility behavior of people. Thus, we consider every day the same way.

\subsection*{Simulation of epidemics}
We use 
the statistical information of Montgomery County, which has approximately 80,000 people, featured by daily activity schedules. There are approximately 430,000 activities and approximately 27,000 distinct locations for the people. 
In the daily simulation, individuals follow the epidemic transmission
rule (see Methods). Different interaction places mean different
transmission rates. Approximately 20\% of people can take public
transit. We simply assume that all 20\% of the people will take
public transit if they can as the initial condition. We employ
different random seeds for the following simulations. The results are
similar. For the following results, we simply choose a random seed randomly. 

\renewcommand{\figurename}{Fig.}

\subsection*{Simulation of interventions}
\begin{figure}[h]
        \centering
        \includegraphics[width=0.80\textwidth]{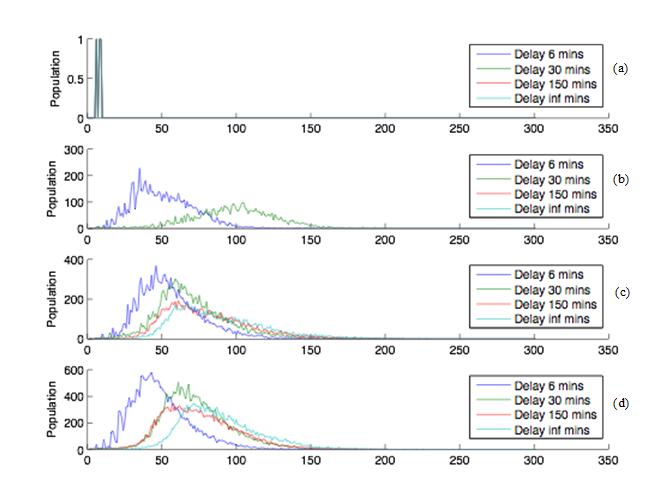}
        \caption{Simulation of the influenza epidemic
          curve with 50\% PTTCR. The number of people who take public transit is
          reduced by 50\%. Additionally, seven scenarios were
          simulated for this intervention. In the four scenarios, the
          time threshold h (see Methods) in public transit is
          increased by 6 minutes, 30 minutes, 150 minutes and an infinite number of
          minutes, and the basic reproductive number R0 is increased by
          (a) R0=0.9, (b) R0=2.1, (c) R0=4.0, and (d) R0=4.5}
          \label{Fig:F3}
    \end{figure}

\begin{figure}[h]
        \centering
        \includegraphics[width=0.80\textwidth]{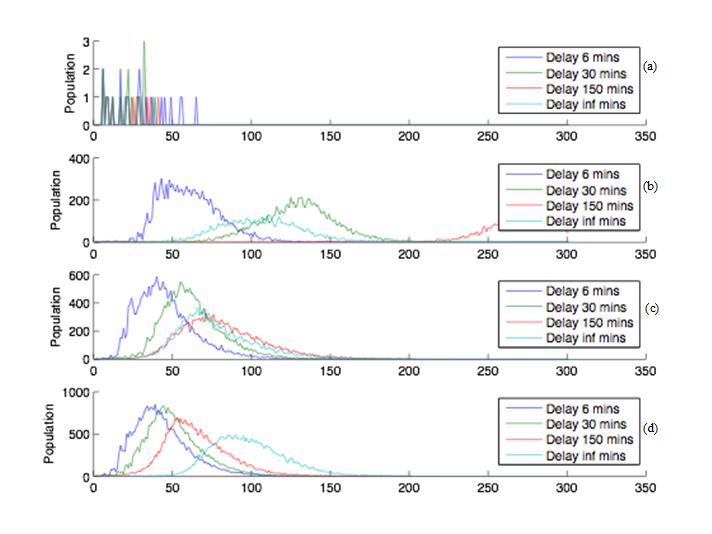}
        \caption{Simulation of the influenza epidemic
          curve with 30\% PTTCR. The number of people who take public transit is
          reduced by 30\%. Additionally, seven scenarios were
          simulated for this intervention. In the four scenarios, the
          time threshold h (see Methods) in public transit is
          increased by 6 minutes, 30 minutes, 150 minutes and an infinite number of
          minutes, and the basic reproductive number R0 is increased by
          (a) R0=0.9, (b) R0=2.1, (c) R0=4.0, and (d) R0=4.5}
          \label{Fig:F4}
    \end{figure}

\begin{figure}[h]
        \centering
        \includegraphics[width=0.80\textwidth]{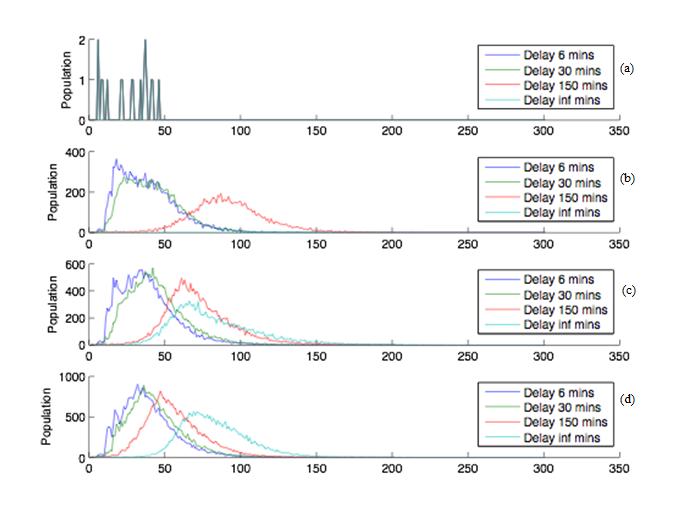}
        \caption{Simulation of the influenza epidemic
          curve with 0\% PTTCR. The number of people who take public transit is
          reduced by 0\%. Additionally, seven scenarios were simulated
          for this intervention. In the four scenarios, the time
          threshold h (see Methods) in public transit is increased by
          6 minutes, 30 minutes, 150 minutes and an infinite number of minutes, and
          the basic reproductive number R0 is increased by (a) R0=0.9,
          (b) R0=2.1, (c) R0=4.0, and (d) R0=4.5}
          \label{Fig:F5}
    \end{figure}

In an epidemic, an individual visiting places and taking public transit 
can impact the transmission of the virus and contact with people,
thereby resulting in dynamic behaviors of the epidemic. Our PTF considers the process. 
To measure the effect of an individual visiting public transit during the epidemic, the following experiments are conducted by setting 
different public transit trip contribution rates in the different
basic reproductive number R0 (see  Figs. \ref{Fig:F3} -
\ref{Fig:F5}). From these three figures, we can see that as the number of people who
take public transit is reduced, less people will be infected. In each
figure, a higher R0 corresponds to a larger infected population. In each subplot,
a larger time threshold h corresponds to a lower infected
population. Moreover, when only 10\% of people take public transit, the infected
population is almost the same for the different h values (30, 150 and inf),
as shown in Fig.~\ref{Fig:F3}. This means that when there are only a
few people who take public transit, the intervention performance of
public transit cannot play a large role in the epidemic
process. Regardless of how much effort is expended here, we cannot
obtain a substantial improvement.

\begin{figure}[h]
        \centering
        \includegraphics[width=0.80\textwidth]{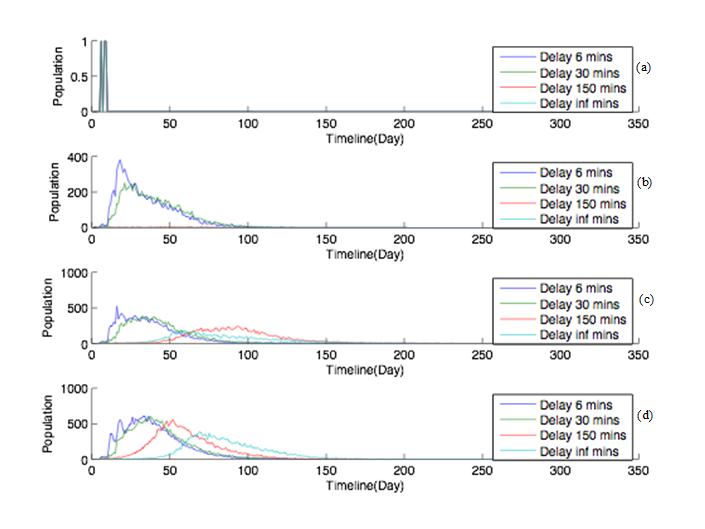}
        \caption{Simulation of the influenza epidemic
          curve with school closure. The time of school closure is 
          considered from the beginning. In the four scenarios, the time
          threshold h (see Methods) in public transit is increased by
          6 minutes, 30 minutes, 150 minutes and an infinite number of minutes, and
          the basic reproductive number R0 is increased by (a) R0=0.9,
          (b) R0=2.1, (c) R0=4.0, and (d) R0=4.5}
          \label{Fig:F2}
    \end{figure}

Rather, when approximately 20\% of people take public transit, the infected
population is different for the different h values (30, 150 and inf), as shown in
Fig.~\ref{Fig:F5}. This means that when there are many people who take
public transit, the intervention performance of public transit will
play an important role in the epidemic process. We can obtain a substantial
improvement if we can control the spreading in public
transit. School closure is a common intervention in an
epidemic. We also simulated two other severe epidemics by closing
schools (see Fig.~\ref{Fig:F2}). Compared with no intervention (see
Fig.~\ref{Fig:F5}), we can see that the effect of the school closure
intervention is more evident when R0 is larger (see
Fig.~\ref{Fig:F2}). The infected population is smaller, and the peak
will arrive later. We can see that the intervention of reducing the public transit trip contribution rate to 10\% has similar and even better performance.

\section*{Conclusion}
An influenza epidemic is caused by many
factors. There is a difference between urban transport in
China and other Western countries; people would like to choose public
transit for their daily travel. 
We used actual data to study pandemic influenza and to explore the
importance of the public transit behavior of the individual. We were able to
investigate the public transit trip contribution rate (PTTCR) in
epidemic spreading processes of the individual and thereby assess the
impact of the public transit trip contribution rate by evaluating the number of infectious people. 
Public transit and school closure strategies were simulated and analyzed.

Our simulation results suggest that individuals with high public
transit trip contribution rates will increase the number of infectious
people when there is an infectious disease outbreak, similar to the school
closure intervention. We conclude that the public transit trip
contribution rates will have an impact on the process of the spread of an infectious
disease because they can affect the social network. In this respect, our work provides a means for modeling the impact of public transit trips and
for estimating the effectiveness of infection controls for public transit trips.


\section*{Supporting Information}
\subsection*{Methods}
\paragraph{Analysis of the transmissibility threshold T\textsubscript{c}}

\begin{equation}\label{eq:a} 
{k}_{i}={k}_{bus,i}+{k}_{\tilde{bus},i}
\end{equation}

where k\textsubscript{bus,i} represents the new degree that the i-th
person receives from the public transit and k\textsubscript{b\~us,i} is the degree that the i-th person receives, even without taking the public transit. Therefore, the average network degree $\langle k \rangle$  is:

\begin{equation}\label{eq:b} 
<k>=\frac{\sum_{i}^{m}{k}_{i}}{m}=\frac{\sum_{i}^{m}{k}_{bus,i}+\sum_{i}^{m}{k}_{\tilde{bus},i}}{m}=<{k}_{bus,i}>+<{k}_{\tilde{bus},i}>
\end{equation}

If we can reduce the average degree  $\langle k \rangle$ to  $\langle k\textsubscript{$\alpha$ } \rangle$ through the intervention of public transit, then:

\begin{equation}\label{eq:c} 
<{k}_{\alpha }>=\alpha <{k}_{bus}>+<{k}_{\tilde{bus}}>
\end{equation}

where $\alpha$ is the remaining percentage of $\langle k\textsubscript{bus} \rangle$ after the intervention of public transit. We assume the following parameters:

\begin{equation}\label{eq:d} 
w=\frac{<{k}_{\alpha }>}{k}=\frac{\alpha <{k}_{bus}>+<{k}_{\tilde{bus}}>}{<{k}_{bus}>+<{k}_{\tilde{bus}}>}=\frac{\alpha \mu +1}{\mu +1}
\end{equation}

\begin{equation}\label{eq:e} 
\mu =\frac{ <{k}_{bus}>}{<{k}_{\tilde{bus}}>}+\frac{1-w}{w-\alpha }
\end{equation}

\begin{equation}\label{eq:f} 
v=\frac{<{{k}_{\alpha }}^{2}>}{<{k}^{2}>}
\end{equation}

Then, the size of the epidemic disease outbreak is:

\begin{equation}\label{eq:g} 
<s>=1+\frac{Tw<k>}{1-T\frac{v<{k}^{2}>-w<k>}{w<k>}}
\end{equation}

The transmissibility threshold T\textsubscript{c} is:

\begin{equation}\label{eq:h} 
\frac{1}{{T}_{c}}=\frac{<{k}^{2}>}{k}-1=\frac{({\alpha }^{2}<{{k}_{bus}}^{2}>)+(<{{k}_{\tilde{bus}}}^{2}>)+(2\alpha <{k}_{bus}{k}_{\tilde{bus}}>)}{\alpha<{k}_{bus}>+<{k}_{\tilde{bus}}>}-1
\end{equation}

Then, we can see that the smaller $\alpha$ is, the lower the average degree
$\langle k \textsubscript{$\alpha $} \rangle$ is and the higher the
threshold T\textsubscript{c} is. This means that the intervention of public transits can affect the trend of epidemic diseases to some extent.

\subsection*{Experiment Simulation}
\paragraph{Public-transit-based modeling framework}

We use a general computational approach for networked epidemiology
based on~\cite{bib14}, which can generate a social contact network of
the region under consideration. Three main steps are involved in the
process of constructing synthetic populations~\cite{bib14}: 
Step 1 constructs an artificial population with open-source databases. 
Step 2 connects daily activities to individuals for each household through daily surveys~\cite{bib14,bib15}.
Step 3 assigns public-transit-based activities between the two activity locations.
Demographics and home locations are considered here. The social contact graph is constructed with  activity information for each person. 
Google Maps can be used here to compute the public-transit-based activities between the two activity locations.

\paragraph{Individual-based model}

\begin{figure}[ht]
        \centering
        \includegraphics[width=0.80\textwidth]{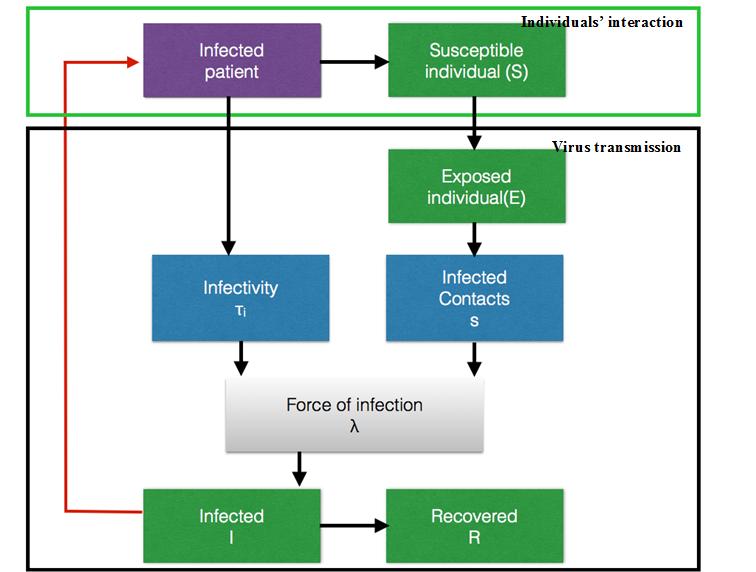}
        \caption{PTF describes individual interaction and virus transmission.}
        \label{Fig:F1}
\end{figure}

We embedded an individual-based model in the PTF model to denote the
virus transmission mechanism by considering the travel of an individual
using public transit (see Fig.~\ref{Fig:F1}). The individual-based model of an epidemic features 
a dynamic process, including factors such as individuals visiting places, visiting public transits, and daily infection transmission.

To model this process, we defined the following functions.
The subject’s infection risk at location p is defined as:
\begin{equation}\label{eq:i} 
\tau (i,j,p)= \beta \cdot c \cdot t(i,j)
\end{equation}
where $\beta$  indicates the infection rate. t(i,j) represents the
contact time between individual i and individual j at location p. c
represents the contact rate and is set as two different values
according to the current location, which belongs to the public
transit set S\textsubscript{public transits} or the place set
S\textsubscript{places}. It is assumed that if someone remained near a
symptomatic patient for more than h hours in the public transit, then the infection rate for this person is 100 \% of ${c}_{p}$, 
where a is the infectiousness at a certain location. If the person
remained at some location for h hours, then the probability decreases proportionally to the duration.

 \begin{equation}
 c=\begin{cases}
\alpha  \cdot{c}_{p} \quad if \quad  p\in {S}_{public\quad transits}\\
{c}_{p}\quad \quad \quad if \quad  p\in {S}_{places}

\end{cases}
\end{equation}

The infection force of susceptible individual i caused by the infected neighbors $j (j=1,…,{S}_{i})$ at location p: $\lambda (i,p) = 1- \prod_{{S}_{i}}^{j=1}(1-\tau (i,j,p))$

where $\tau (i,j,p)$ is the infectivity of infected contact j at location p, capturing the probability of infected individual j infecting others.

\subsection*{Acknowledgments}
 This work was supported by the National Natural Science Foundation of
 China 61272412. The funders had no role in the study design, data collection and analysis, decision to publish, or preparation of the manuscript.



%
%
%

\end{document}